\documentstyle [12pt] {article}

\input epsf

\newcommand{\be}{\begin{equation}}
\newcommand{\ee}{\end{equation}}
\newcommand{\beqs}{\begin{eqnarray}}
\newcommand{\eeqs}{\end{eqnarray}}
\def\({\left (}
\def\){\right )}
\def\l[{\left[}
\def\r]{\right]}
\def\zlZ{{\zeta_2 \over \zeta_1}}
\def\zZl{{\zeta_1 \over \zeta_2}}
\def\zl{\zeta_1}
\def\zZ{\zeta_2}
\def\z{\zeta}
\def\d{\delta}

\def\L{\Lambda}
\def\a{\alpha}

\def\na{\nabla}

\def\da{\dot{\alpha}}

\def\pa{\partial}

\def\U{\Upsilon}

\def\bU{\bar{\Upsilon}}

\def\th{\theta}
\def\ch{{\cal H}}
\def\ni{\noindent}
\def\nn{\nonumber}

\begin{document}

\begin{titlepage}

\begin{flushright}
\begin{tabular}{l}
ITP-SB-98-25    \\
hep-th/9804010  \\ 
April, 1998 
\end{tabular}
\end{flushright}

\vspace{8mm}
\begin{center}
{\Large \bf Nonholomorphic N=2 terms in N=4 SYM: }
 
\medskip
{\large \bf 1-Loop  Calculation in N=2 superspace}

\vspace{4mm}
\vspace{16mm}

F. Gonzalez-Rey \footnote{email: glezrey@insti.physics.sunysb.edu} and 
M. Ro\v{c}ek \footnote{email: rocek@insti.physics.sunysb.edu}

\vspace{4mm}
Institute for Theoretical Physics  \\
State University of New York       \\
Stony Brook, N. Y. 11794-3840  \\

\vspace{20mm}

{\bf Abstract}
\end{center}

 The effective action of $N=2$ gauge multiplets in general includes 
higher-dimension UV finite nonholomorphic corrections integrated with 
the full $N=2$ superspace measure. By adding a hypermultiplet in the 
adjoint representation we study the effective action of $N=4$ SYM. 
The nonanomalous $SU(4)$ R-symmetry of the classical $N=4$ theory must
be also present in the on-shell effective action, and therefore we 
expect to find similar nonholomorphic terms for each of the scalars 
in the hypermultiplet. 
The $N=2$ path integral quantization formalism developed in projective 
superspace allows us to compute these hypermultiplet nonholomorphic terms 
directly in $N=2$ superspace. The corresponding gauge multiplet 
expression can be successfully compared with the result inferred from a 
$N=1$ calculation in the abelian subsector.

\end{titlepage}
\newpage
\setcounter{page}{1}
\pagestyle{plain}
\pagenumbering{arabic}
\renewcommand{\thefootnote}{\arabic{footnote}}
\setcounter{footnote}{0}

\section{Introduction}

 The nonholomorphic $N=2$ potential $\ch(W,\bar{W})$ in the 
effective action of SYM theories has been an object of study for some 
time now \cite{non_hol}-\cite{dine_seiberg}. This potential is integrated 
with the full $N=2$ superspace superspace measure and therefore is a
dimensionless real function of the $N=2$ gauge field strengths $W$ and
$\bar{W}$. In the nonabelian sector it contributes to the $N=1$ 
K\"{a}hler potential ${\cal K} (\phi, \bar{\phi})$ and in the abelian 
sector it can only contribute to $N=1$ higher derivative terms 
\cite{n1}. Scale invariance and $U(1)_R$ invariance restricts the form
of the $N=2$ potential to be 

\be
 \ch = \ch^o + c \ln {W^2 \over \L^2 } \ln {\bar{W}^2 \over \L^2 } \ ,
\label{nonhol_kahler}
\ee

\ni
where $\ch^o$ depends on gauge invariant, scale independent
combinations of the nonabelian $N=2$ field strengths. The pure abelian
piece is contained only in the second term. In $N=4$ SYM theories the
abelian nonholomorphic potential is believed to be generated only at 
1-loop \cite{dine_seiberg} since higher loop and nonperturbative
contributions would break the scale and $U(1)_R$ invariance of $\ch$.
Nonperturbative contributions have been studied in \cite{dorey} 
and they give vanishing results.

 It is therefore possible to determine the exact form of $\ch^{Abelian}$
by performing a 1-loop calculation. Recently this type of calculation 
has been done in $N=1$ superspace \cite{rikard_vipul} by considering 
the higher derivative operator

\be
 \int d x d^4 \th \( i \ch_{A \bar{B}} \bar{W}^{B \da} \na^\a_{\da} 
 W^A_\a \)  \ .
\label{n1_nonhol}
\ee

\ni
This is one of the $N=1$ components of (\ref{nonhol_kahler}) \cite{n1}.  
Its contribution to the 1-loop abelian effective action was computed 
using $N=1$ superspace quantization. The resulting coefficient $c$ was 
found to be nonvanishing. This has interesting implications for 3-branes
in ten dimensions because they are believed to be effectively described
by $N=4$ SYM at low energies: the presence of a nonvanishing $N=2$
nonholomorphic potential introduces acceleration dependent terms in
the scattering of 3-branes in addition to the standard velocity 
dependent terms \cite{rikard_vipul}.

 In this article we compute the nonholomorphic corrections to the
$N=4$ SYM effective action directly in $N=2$ superspace using the
$N=2$ path integral quantization that we developed in \cite{n2_hyper},
\cite{n2_gauge}. This quantization involves $N=2$ superfields that 
contain the familiar $N=1$ hypermultiplet and gauge degrees of 
freedom\footnote{Related but different $N=2$ superfield Feynman rules
have been developed in harmonic superspace \cite{gal}. Up to now
they have not correctly reproduced the calculations we 
describe here \cite{disagree} .}.

 First we calculate all finite 1-loop corrections to
the $N=2$ hypermultiplet effective action dropping terms with spinor
or space-time derivatives on the external fields. The calculation is
greatly simplified using $N=2$ gauge propagators in the Landau gauge.

 We then isolate the dependence of the effective action 
on the $N=2$ hypermultiplet superfield whose $N=1$ projection is part  
of the chiral hypermultiplet isodoublet $\bU_0 |_{\th^2_\a = 0} 
= \tilde{Q}$. This contribution to the $N=2$ effective action is a 
nonholomorphic potential $\ch(\bU_0, \U_0)$ whose $N=1$ projection 
can be rotated by a $Z_2$ subgroup of the global $SU(4)_R$ of $N=4$ 
SYM into the $N=1$ projection of a corresponding pure gauge piece 
$\ch(W,\bar{W})$. 
Symmetry arguments therefore determine the form of the 1-loop $N=2$
nonholomorphic potential in the low energy gauge effective action of 
$N=4$ SYM.

 The $N=2$ potential we find for the abelian sector is of the form
(\ref{nonhol_kahler}). The coefficient $c$ is exactly the same as that
calculated in $N=1$ superspace \cite{rikard_vipul}. Due to the
nonlinearity of the nonabelian superfield strengths it is not clear 
if the 1-loop nonabelian piece $\ch^o$ can be reproduced from the 
knowledge of the hypermultiplet effective action and we cannot test the 
proposal in \cite{n1}.

\section{N=2 formalism}

 In this section we briefly review the superfield content of the  
$N=4$ SYM in $N=2$ superspace and we give the Feynman rules for 
quantization of these $N=2$ multiplets. For a more detailed explanation
we refer the reader to \cite{n2_hyper}-\cite{n2_gauge} and references 
therein. The conventions we follow are those of ref. \cite{book}.

 Gauge multiplets and hypermultiplets can be described by off-shell 
representations of $N=2$ supersymmetry using superfields that live in 
projective superspace. This is a subspace of $N=2$ superspace whose 
anticommuting coordinates are the following linear combinations of the $N=2$ 
Grassmann coordinates: $\Theta^\a = \th^{2 \a} - \z \th^{1 \a}$ and 
$\bar{\Theta}^{\da} = \bar{\th}_1^{\da} + \z \bar{\th}_2^{\da}$
parameterized by a complex projective coordinate $\z$. Accordingly, 
projective superfields $\Omega$ obey the constraint

\be
 \na_\a \Omega (\Theta, \bar{\Theta}) = (D_{1\a} + \z D_{2\a}) \Omega = 0 
 = (\bar{D}^2_{\da} - \z \bar{D}^1_{\da} ) \Omega = \bar{\na}_{\da} \Omega
\label{proj_const}
\ee   

 Charged hypermultiplets can be described by an infinite power series in
the projective coordinate\footnote{Under conjugation $\z
\longrightarrow -\z^{-1}$. See \cite{n2_hyper}. }. We refer to this
multiplet as the polar multiplet 

\beqs
 \U & = & \sum_{n=0}^\infty \U_n \zeta^n   \\
 \bU & = & \sum_{n=0}^\infty \bU_n 
  (- {1 \over \zeta})^n \nn  \ . 
\eeqs

 As a consequence of the constraints (\ref{proj_const}) the highest
order coefficient is a chiral superfield in $N=1$ superspace 
$\bar{D}_{\da} \bU_0 = 0$ and the next order is a complex 
linear superfield $\bar{D}^2 \bU_1 = 0$. These two superfields contain
the physical degrees of freedom of the hypermultiplet. The other 
coefficients are auxiliary superfields in $N=1$ superspace.  

 Gauge vector multiplets are described by an infinite series with 
negative and positive powers of 
the projective complex coordinate that we call the tropical multiplet. 
This multiplet is real under conjugation and since there are no lowest
order or highest order coefficients, all of them are unconstrained in
$N=1$ superspace 

\be
 V = \sum_{n= -\infty}^{+\infty} v_n \zeta^n \; , \;\;\; v_{-n} =
 (-)^n \bar{v}_n \ .
\ee

 The coefficients $v_n ,\; |n|>1$ are gauge degrees of freedom, $v_0$ is
related to the usual $N=1$ real gauge prepotential $v= v_0 +
nonlinear \ corrections$ of the covariantly chiral spinor field
strength and $v_{-1}$ is related to the prepotential $\bar{\psi} = i v_{-1}
+ nonlinear \ corrections$ of the ordinary chiral scalar $\phi$ that 
appears in the $N=1$ components of the classical gauge action

\beqs
 S^{gauge} & = & \int dx d^2 \th d^2 \bar{\th} \; {Tr \over 4 g^2} \; 
  e^{-v} \bar{\phi} \, e^{v} \phi \: + {1 \over 2} 
 \( \int d x d^2 \th \; {Tr \over 4 g^2} \; {W^\a W_\a \over 2} + 
 \int d x d^2 \bar{\th} \; {Tr \over 4 g^2}  \; 
 { \bar{W}^{\da} \bar{W}_{\da} \over 2} \) 
\label{n1_gauge_act} \nn  \\
            \nn \\
& & W_\a = i \bar{D}^2 e^{-v} D_\a e^{v} \; , \;\;\; \phi =
 \bar{D}^2 \bar{\psi}  \ .
\eeqs

The manifestly $N=2$ supersymmetric action describing the coupling of
a polar hypermultiplet in the adjoint representation of the gauge group
to the tropical gauge multiplet is the following

\be 
 S_\U =  \int d x d^2 \th d^2 \bar{\th} \oint {d \z \over 2 \pi i \z} 
 \; \; {Tr \over 4 g^2} \(e^{- V} \U \, e^{V} \, \bU  \)  \ .
\label{eq-Lagrangian2}
\ee
 
 This action can be dualized to the give the usual description of the 
hypermultiplet in terms of two chiral fields \cite{n2_gauge}: the 
complex antilinear superfield $\U_1$ is traded for a chiral Lagrange 
multiplier $Q$, the chiral coefficient superfield $\bU_0$ is
identified with its $N=2$ partner $\tilde{Q}$ and the auxiliary
superfields decouple. The resulting interacting action is the usual
one

\be
S_{Q \tilde{Q}} = \int dx d^4 \th \; {Tr \over 4 g^2} 
 \( e^{- v} \bar{\tilde{Q}} \, e^{v} \tilde{Q} + 
 e^{-v} \bar{Q} \, e^{v} Q \) \: + \( i \int dx d^2 \th \; {Tr \over 4 g^2}  
 \; \tilde{Q} [\phi, Q] + h.c. \) \ .
\ee

\ni
The convention we adopt to define the path integral is such that the
kinetic term of scalars is convergent in euclidean space

\be
 Z = \int [d Q] [d \tilde {Q}]  e^{- i S} = \int [d Q] [d \tilde {Q}]
 e^{ S_E} \; , \;\;\; S_E = \int d^4 k_E \: Q(k_E) \: (- k_E^2) \:
 \bar{Q}(-k_E) + \dots \ .  
\ee
 
 For the gauge multiplet, the kinetic piece of (\ref{n1_gauge_act})
has a simple expression in terms of tropical multiplets 

\be
 S_0^{gauge} = - {Tr \over 8 g^2} \int d x d^8 \th \oint 
 {d \zl \over 2 \pi i} {d\zZ \over 2 \pi i} \: { V(\zl) V(\zZ) 
 \over (\zl - \zZ)^2 } \ ,
\label{n2_vector_action}
\ee

\ni
while the interaction vertices are more complicated. Since we are 
only going to compute 1-loop amplitudes with external hypermultiplets 
coupling to internal gauge multiplets, all we need is the gauge 
propagator in projective superspace. The kinetic action 
(\ref{n2_vector_action}) is therefore enough to use the path 
integral quantization of the model.  

 We recall examine the Feynman rules that we use to compute the set
of diagrams proposed. The polar hypermultiplet propagator
is \cite{n2_hyper}

\beqs 
 \langle \bar{\U}^a (1) \; \U^b (2) \rangle & = & 
 - {4 g^2 \, \d^{ab} \over C_2(A)} \sum_{n=0}^{\infty} \(\zlZ\)^n  
 {\na_1^4 \na_2^4 \over \zl^2 (\zl -\zZ)^2 \Box} \, \d^8( \th_1 - \th_2 )
 \, \d (x_1-x_2) \nn \\  
                    & & \nn \\
 \langle \U^a (1) \; \bar{\U}^b (2) \rangle & = & 
 - { 4 g^2 \, \d^{ab} \over C_2(A)} \sum_{n=0}^{\infty} \( {\zZl} \)^n \;
 \; { \na_1^4 \na_2^4 \over \zZ^2 (\zl-\zZ)^2 \Box } \, \d^8 (\th_1 - \th_2) 
 \, \d (x_1-x_2) \ . 
\label{n2_propupsi21}
\eeqs

\ni
where $C_2(A)$ is the second Casimir in the adjoint representation of 
the gauge group $Tr T_a T_b = C_2(A) \: \d_{ab}$ and $\na_1^4 = \na^2 (\zl) 
\bar{\na}^2 (\zl)$. The tropical multiplet propagator is \cite{n2_gauge}

\be
 \langle V^a(1) V^b(2) \rangle = {4 g^2 \, \d^{ab} \over C_2(A)} 
 \( \a \sum_{n=-\infty}^{+\infty} \( \zlZ \)^n + (1 - \a) \) 
 {\na_1^4 \na_2^4 \over \zl \zZ \, (\zZ - \zl)^2 \Box^2 } 
 \, \d^8 (\th_{12}) \, \d^4 (x_{12}) \ ,
\ee

\ni 
where $\a$ denotes the gauge fixing parameter (the factor $g^2$ in the 
propagators is the one consistent with the holomorphic normalization 
of the kinetic term as opposed to canonical normalization in
the sense of \cite{murayama}). The interaction
vertices that contribute to the relevant graphs are obtained from the 
first two orders in the expansion of (\ref{eq-Lagrangian2})

\be
 - \int dx d^4 \th \oint {d \z \over 2 \pi i \z} \; {Tr \over 4 g^2} \:
 [ V, \U] \, \bU \;\;\; + \;\; \int dx d^4 \th \oint {d \z \over 2 \pi i \z}  
 \; {Tr \over 4 g^2} \: {1 \over 2} \: [V,[ V, \U]] \, \bU \nn \ .
\ee

 Now we can use all the powerful tools of path integral quantization
to calculate the 1-loop effective action of the polar multiplet. The
only peculiarity of the $N=2$ formalism is that we must
complete the Grassmann measure of each vertex to have a full $N=2$
superspace measure. This procedure eliminates four projective spinor
derivatives in one of the propagators stemming from the vertex. For 
example in the vertex with an external arctic multiplet

\[
 \int dx D^2 \bar{D}^2 \oint {d \zl \over 2 \pi i \zl} {C_2(A) \over 4 g^2}
 (-i) f_{abc} \U^b (\zl) \l[ {4 g^2 \d^{ad} \over C_2(A)} 
 {\na_1^4 \na_2^4 \, \d_{12} \over \zl \zZ (\zZ - \zl)^2 \Box^2 } \r] 
 \l[ {4 g^2 \d^{ce} \over C_2(A)} {\na_1^4 \na_3^4 \, \d_{13} \over 
 \zl^2 (\z_3 - \zl)^2 \Box } \dots \r]  \nn 
\]

\be
= \int dx D_1^2 (\bar{D^1})^2 D_2^2 (\bar{D^2})^2  
 \oint {d \zl \over 2 \pi i \zl} \U_{ca} (\zl) 
 \l[ { 4 g^2 \d^{ad} \over C_2(A)} {\na_1^4 \na_2^4 \, \d_{12} \over 
 \zl \zZ (\zZ - \zl)^2 \Box^2 } \r] \l[ \d^{ce}  
 { \na_3^4 \, \d_{13} \over (\z_3 - \zl)^2 \Box } \dots \r] 
\ee

 Once the measure has been completed in all the vertices, we do the
``D''-algebra to reduce all propagators but one to bare Grassmann delta
functions. These are the basic Feynman rules that we use in the next
section to construct the 1-loop effective action of the polar multiplet.

\section{1-loop nonholomorphic terms in the hypermultiplet effective action}
 
 Now that we have presented the rules to calculate Feynman diagrams in
$N=2$ superspace, we focus our attention on those amplitudes of
interest to us. We want to consider graphs with any number of external
polar multiplets at zero momentum. The calculation is greatly
simplified working with the gauge propagator in the Landau gauge
$\a=0$. To illustrate the techniques used in this novel
$N=2$ quantization we present the simplest graphs in some detail. The
more complicated ones only involve a larger amount of algebra.

\begin{figure}[hbt]
\begin{center}
\mbox{\epsfysize=5.56cm \epsfxsize=9cm \epsfbox[60 150 510 428]{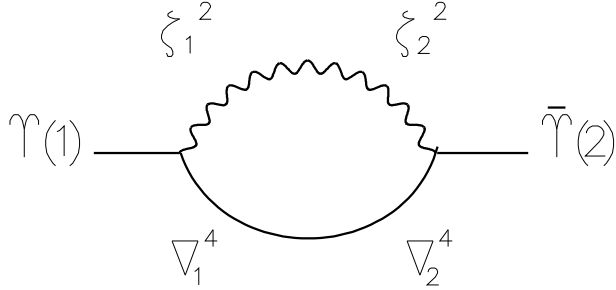}}
\caption{Hypermultiplet 2-point function.}
\end{center}
\label{2pt_hyp}
\end{figure}

In the $N=2$ formalism tadpoles and seagulls
vanish automatically \cite{n2_hyper}, and at one loop we always have 
the same number of external arctic and antarctic polar fields. Therefore 
the first graph we study is the two point function. After completing 
the Grassmann measure on both vertices we find the graph on 
Fig. \ref{2pt_hyp} and a similar graph in which the external 
hypermultiplets are exchanged.

 The ``D''-algebra on this graph is trivial: we already have one 
bare propagator and the other one is acted upon by eight 
spinor derivatives. We just have to reduce

\be
 \d^8 (\th_1-\th_2) \na_1^4 \na_2^4 \: \d^8 (\th_1-\th_2) = (\zl-\zZ)^4 
 \: \d^8 (\th_1-\th_2) \ .
\ee

\ni
 
The resulting contribution to the effective action $-i \Gamma (2)$ is
UV finite (this is the well known nonrenormalization of
hypermultiplets) and local in the $N=2$ Grassmann coordinates

\be     
 {1 \over 2!} \oint {d \zl \over 2 \pi i \zl} 
 {d \zZ \over 2 \pi i \zZ} \int d^8 \th {i d^4 p \over (2 \pi)^4} 
 {-1 \over \(-p^2 \)^3 } \: Tr \( \sum_{n=1}^{\infty} \(\zlZ\)^n
 \U (\zl) \bU (\zZ) + \sum_{n=1}^{\infty} \(\zl \over \zZ \)^n
 \bU (\zl) \U (\zZ) \)   \ .
\label{graph_1}
\ee

\begin{figure}[htp]
\begin{center}
\mbox{\epsfysize=4.5cm \epsfxsize=9.5cm \epsfbox[90 190 470 370]
{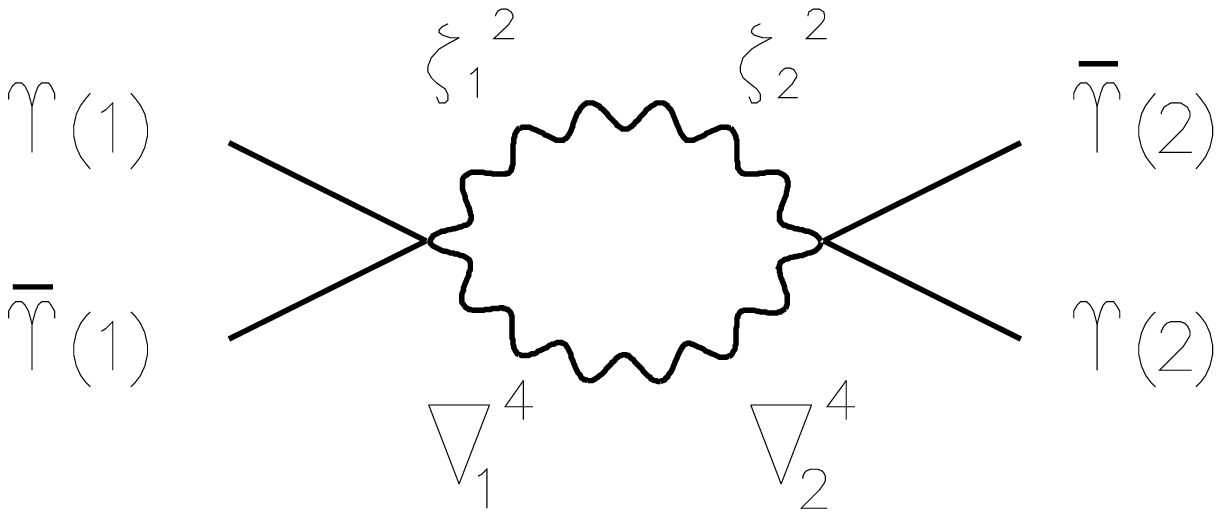}}
\end{center}
\begin{center}
\mbox{\epsfysize=5.75cm \epsfxsize=9.5cm \epsfbox[90 175 470 405]
{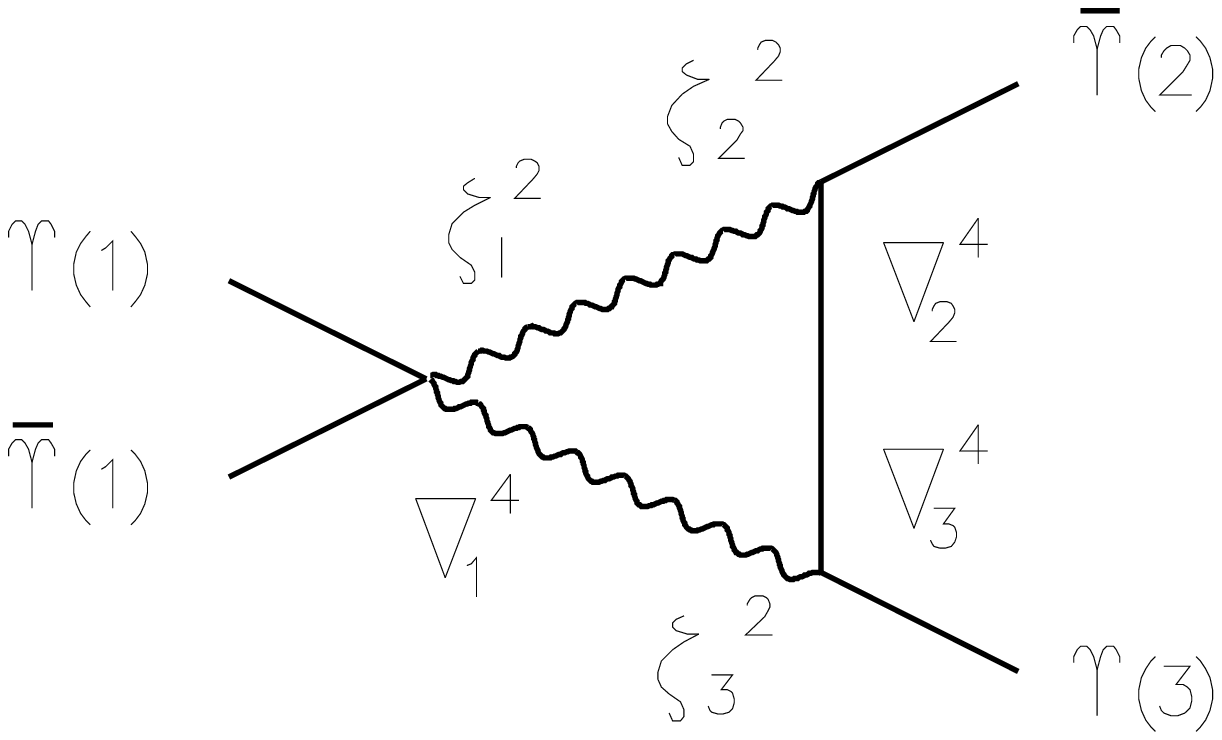}}
\end{center}
\begin{center}
\mbox{\epsfysize=6.1cm \epsfxsize=8.2cm \epsfbox[90 135 500 440]
{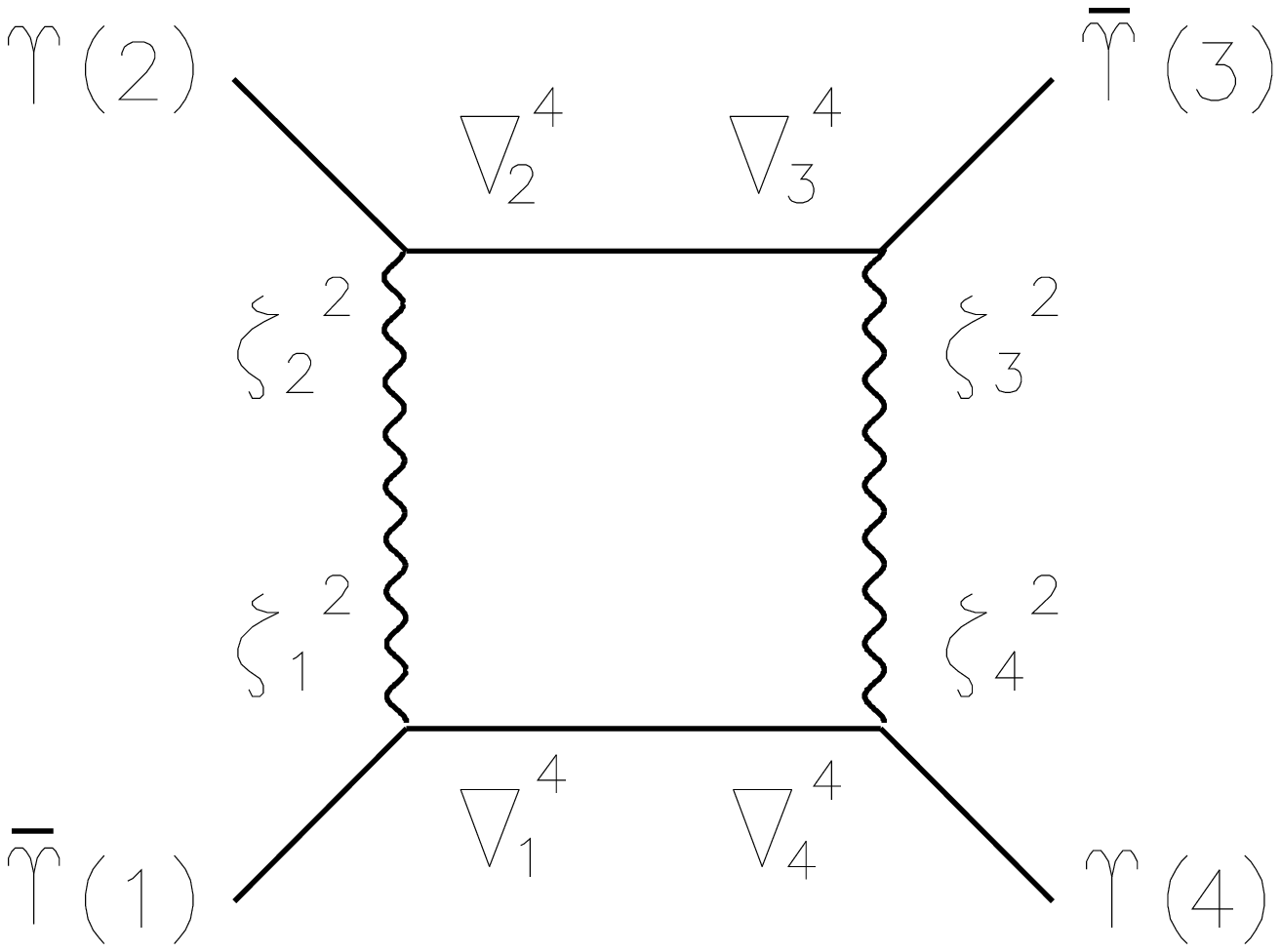}}
\end{center} 
\caption{Hypermultiplet 4-point function.}
\label{4pt_hyp}
\end{figure}

Next we consider the graphs with four external polar multiplets. After
completing the superspace measure in the vertices we obtain the graphs in 
Fig. \ref{4pt_hyp} and similar ones in which we exchange the external
hypermultiplet of each cubic vertex by the internal hypermultiplet.

 The ``D'' algebra is trivial only on the upper graph. In the other two
we integrate by parts the spinorial derivatives of all
propagators but one. As usual, it is easiest to do so on the graph.
Since we are interested on nonholomorphic terms without external
derivatives we keep only the contribution where all spinor
derivatives end up acting on the same propagator.

 Finally we can integrate the bare Grassmann delta functions and reduce
the spinor derivatives acting on the last one. For example in the 
middle graph

\beqs
\lefteqn{\d^8 (\th_1-\th_2) \na_3^2 \bar{\na}_3^2 \na_1^2 \bar{\na}_1^2 
 \bar{\na}_2^2 \na_2^2 \d^8 (\th_1-\th_2) = (\z_3 - \zl)^2 \Box \,
 \d^8 (\th_1-\th_2) \na_3^2 \bar{\na}_1^2 \bar{\na}_2^2 \na_2^2 
 \, \d^8 (\th_1-\th_2)      \nn } \\
& & \qquad \qquad \qquad \qquad \qquad \qquad \qquad  
= (\z_3 - \zl)^2 (\z_3 - \zZ)^2 (\zl - \zZ)^2 \Box \, 
 \d^8 (\th_1-\th_2) \ . 
\eeqs

 As a result all the graphs with four external hypermultiplets and no
external derivatives have the same euclidean loop momentum integral 
$\int i d^4 p/(- 2 \pi p^2)^4$. The complex coordinate dependence
is slightly different though. The first graph gives a contribution to 
$-i \Gamma (4)$

\be
 {1 \over 2!} \oint {d \zl \over 2 \pi i \zl} {d \zZ \over 2 \pi i \zZ}  
 \: Tr {1 \over 4} \l[ \( \U (\zl) \bU (\zl) + \bU (\zl) \U (\zl) \)  
 {\U (\zZ) \bU (\zZ) + \bU (\zZ) \U (\zZ) \over 2} \r] \ ,
\ee  

\ni
the second is

\beqs
 - {1 \over 2} \oint {d \zl \over 2 \pi i \zl} 
 {d \zZ \over 2 \pi i \zZ} {d \z_3 \over 2 \pi i \z_3} & Tr & 
 \l[ {\U (\zl) \bU (\zl) + \bU (\zl) \U (\zl) \over 2} \right. \\
& & \left. \times \( \sum_{n=1}^{\infty} \(\z_3 \over \zZ \)^n 
 \U (\zZ) \bU (\z_3) + \sum_{n=1}^{\infty} \(\zZ \over \z_3 \)^n 
 \bU (\zZ) \U (\z_3) \) \r] \nn 
\eeqs

\ni
and the last one  

\beqs
 {1 \over 4} \oint {d \zl \over 2 \pi i \zl} 
 {d \zZ \over 2 \pi i \zZ} {d \z_3 \over 2 \pi i \z_3} 
 {d \z_4 \over 2 \pi i \z_4} & Tr & \l[
 \( \sum_{n=1}^{\infty} \(\zZ \over \z_3 \)^n \U (\z_3) \bU (\zZ)
 + \sum_{n=1}^{\infty} \(\z_3 \over \zl\)^n \bU (\z_3) \U (\zZ) \) 
                 \right. \\ 
& & \left. \times \( \sum_{n=1}^{\infty} \(\zl \over \z_4 \)^n 
 \U (\z_4) \bU (\zl) + \sum_{n=1}^{\infty} \(\z_4 \over \zl \)^n 
 \bU (\z_4) \U (\zl) \) \r] \nn
\eeqs

 This simple result illustrates a few features that will be reproduced by 
higher n-point functions: 
 
\begin{itemize} 
 \item[i)] integrating the complex coordinates we can see that the
           coefficient superfields enter quadratically $\U_i \bU_i$; 
 \item[i)] graphs containing one or more internal hypermultiplet
           propagators do not contribute terms that depend purely on 
           powers of $\U_0 \bU_0$.  
\end{itemize}

 This simplifies our calculation considerably because we are interested
in selecting terms that depend only on the $N=2$ superfield containing
$\tilde{\bar{Q}} = \U_0|_{\th^2_\a = 0}$. Terms mixing auxiliary 
superfields $\U_i , \ i>1$ and $\U_0$ do not modify the pure 
$\bU_0 \U_0$ piece because they enter at least quadratically. We may 
set the auxiliary fields to zero using their algebraic field equations.

 We focus our attention on $\U_0$ because the $N=1$ superfield 
$\tilde{\bar{Q}}$ is rotated by $Z_2$ subgroup of $SU(4)_R$ into the 
$N=1$ gauge scalar $\bar{\phi}$. Since this symmetry is nonanomalous 
the nonholomorphic potential

\be
 \int d^8 \th \; \ch (\U_0, \bU_0) 
\label{propos}
\ee
 
\ni
must be accompanied by a corresponding nonholomorphic function of  
$N=2$ superfields whose $N=1$ projection is precisely $\bar{\phi}, \phi$.  
These are the $N=2$ chiral gauge field strengths $W, \bar{W}$.

 Now that we know what kind of amplitudes to look for, we
collect all the relevant graphs with any number of external
hypermultiplets but no internal hypermultiplets and find

\be
 \sum_{m \geq 4} \Gamma (m)= - {1 \over 2} \int d^8 \th \: 
 {d p^2 \over (4 \pi)^2 \, p^2} \sum_{n=2}^{+\infty} 
 Tr \, {1 \over n} \( {\U_0 \bar{\U}_0 + 
 \bar{\U}_0 \U_0 \over - 2 \, p^2} \)^n \ .
\label{relevant}
\ee

 This is almost the Taylor expansion of a logarithm but it is missing
the first order term. To find this term let us go back for a moment to
the result of the first graph (\ref{graph_1}). It does not seem to
contain a piece depending on the chiral superfield we are interested
in $\bU_0 \U_0$. Notice however that it is possible to rewrite the 
contour integrals in (\ref{graph_1}) as follows

\be
 \oint {d \zl \over 2 \pi i \zl} {d \zZ \over 2 \pi i \zZ} 
 \sum_{n=1}^{\infty} \(\zl \over \zZ \)^n Tr \, \U (\zl) \bU (\zZ) = 
 \oint {d \z \over 2 \pi i \z} \, Tr \( \U (\z) \bU (\z) - \U_0 \bU_0 \) \ . 
\ee

\ni
The first term is a projective quantity and therefore it vanishes if
we integrate it with the full $N=2$ superspace measure. Thus we obtain
a contribution to the effective action that we can write in two
equivalent ways 

\be
 \Gamma(2) = \int {d p^2 \over (4 \pi)^2 p^2} \, 
 Tr  \sum_{i>1} {\U_i \bU_i + \bU_i \U_i \over - 2 \, p^2 } = 
 - \int {d p^2 \over (4 \pi)^2 p^2} \,
 Tr \, {\U_0 \bU_0 + \bU_0 \U_0 \over - 2 \, p^2 } \ .
\ee

 To help us decide which form we use let us recall that the physical 
superfield $\U_1|_{\th^2_\a=0}$ is mapped by duality to one of the 
chiral fields $Q$ of the hypermultiplet on-shell description and the 
superfield $\bU_0|_{\th^2_\a=0}$ 
is identified with its partner $\tilde{Q}$. Since we expect the global
$SU(2)_R$ symmetry of these two chiral fields to be realized in the
effective action\footnote{Actually in $N=4$ SYM this is just a
subgroup of the larger $SU(4)_R$ we mentioned before.} it seems natural to
choose an equally weighted combination 
 
\be 
 - {1 \over 4 p^2} \: Tr \sum_{i>1} (\U_i \bU_i + \bU_i \U_i) + 
 {1 \over 4 p^2} \: Tr \, (\U_0 \bU_0 + \bU_0 \U_0) \ .
\label{equipartition}
\ee

 This choice will prove to be correct when we compare the corresponding 
nonholomorphic gauge effective action with the result inferred from 
its $N=1$ components \cite{rikard_vipul}.

Adding (\ref{relevant}) and the $\U_0 \bU_0$ piece in (\ref{equipartition})
we find the nonholomorphic contribution to the effective action

\be
 \int d^8 \th \, \ch (\U_0, \bU_0) = {1 \over 2} \int d^8 \th \, 
 {d p^2 \over (4 \pi)^2 p^2} \, Tr \ln \( 1 + {\U_0 \bar{\U}_0 + 
 \bar{\U}_0 \U_0 \over 2 p^2} \) \ .
\ee

\ni
The corresponding nonholomorphic potential for the $N=2$ gauge field 
strength $W$ is therefore

\be
 \ch (W, \bar{W}) = {1 \over 2} \int {d p^2 \over (4 \pi)^2 
 p^2} \, Tr \ln \( 1 + {W \bar{W} + \bar{W} W \over 2 p^2} \) \ . 
\label{n2_result}
\ee

To simplify our analysis let us consider the case of $SU(2)$ SYM. The 
gauge operator in the argument of the logarithm can be diagonalized 
\cite{n1} 

\be
 U \, (W \bar{W} + \bar{W} W) \, U^\dag = 
 \( \begin{array}{ccc}
          2 W \cdot \bar{W} & 0 & 0 \\
           0 & W \cdot \bar{W} + \sqrt{W^2 \bar{W}^2} & 0 \\     
           0 & 0 & W \cdot \bar{W} - \sqrt{W^2 \bar{W}^2} 
       \end{array} \) \ .
\label{eigen_su2}
\ee

\ni
This facilitates the evaluation of the trace in (\ref{n2_result})

\beqs
 \ch (W, \bar{W}) & = & {1 \over 2 (4 \pi)^2} \int {d p^2 \over p^2} 
 \; \ln \( 1 + {W \cdot \bar{W} \over p^2} \) + 
 \ln \( 1 + {W \cdot \bar{W} + \sqrt{W^2 \bar{W}^2}  \over 2 p^2} \) 
 \nn \\ 
& & \qquad \qquad \qquad + 
 \ln \( 1 + {W \cdot \bar{W} - \sqrt{W^2 \bar{W}^2}  \over 2 p^2} \) \ . 
\label{h-1loop}
\eeqs

\section{Loop momentum integration and regularization issues}

 We have obtained the 1-loop N=2 nonholomorphic potential of $N=4$ SYM
as a loop momentum integral. Now we want to study the abelian sector
of the theory by performing this momentum integral.

Then we can compare our result with the 
explicit form (\ref{nonhol_kahler}) consistent with scale and $U(1)_R$
invariance. We can also compare the coefficient $c$ obtained in $N=1$ 
superspace \cite{rikard_vipul} with the coefficient we obtain. Let us 
briefly review the result of the $N=1$ calculation. The sum of all 1-loop
amplitudes with two external abelian spinor field strengths , one space 
time derivative and arbitrary numbers of abelian gauge scalars is of the 
form (\ref{n1_nonhol}) with

\be 
 {\pa^2 \over \pa W \pa \bar{W} } \ch^{Abel} 
 (W, \bar{W})\left|_{\th^2_\a =0} \right.  = 
 {1 \over (4 \pi)^2 } {1 \over \phi \bar{\phi} } \ .  
\label{n1_result}
\ee

\ni
Promoting the $N=1$ chiral field strengths to $N=2$ chiral
field strengths it is straightforward to integrate $\ch_{W \bar{W}}$ 
with respect to $W$ and $\bar{W}$ to obtain the postulated result

\be
 \ch^{Abel} = {1 \over (4 \pi)^2 } \ln W \ln \bar{W} \ .
\ee

 Let us consider now the abelian piece of (\ref{h-1loop}). When $W$
and $\bar{W}$ commute the third term vanishes and the other two give 
equal contributions

\be
 \ch^{Abel} (W, \bar{W}) = 2 \times {1 \over 2 (4 \pi)^2} 
 \int {d p^2 \over p^2} \ln \( 1 + {W \bar{W} \over p^2} \) \ .
\label{my_result}
\ee

\ni
Now we have to perform the loop momentum integral and verify that we 
reproduce the postulated form (\ref{nonhol_kahler}) of $\ch^{abel}$. 
To show that the dependence on any mass scale is irrelevant we
regulate the divergences of this integral in two different ways. First
we introduce an IR cutoff $\L^2$ and we rescale the loop momentum into a
dimensionless variable   

\be
 \ch^{Abel} = {1 \over (4 \pi)^2} 
 \int_{\L^2 \over W \bar{W}}^{\infty} {d y \over y} \ln \, (1 + y^{-1}) 
 = {1 \over (4 \pi)^2} \int_{0}^{W \bar{W} \over \L^2} 
 \, {d x \over x} \ln (1 + x) \ .  
\ee

The lower limit of the integral does not give any contribution. 
To see this we split the integral in two pieces

\be 
 \int d^8 \th \: \ch^{Abel} = {1 \over (4 \pi)^2} \int d^8 \th
 \( \int_0^{\xi} + 
   \int_{\xi}^{W \bar{W} \over \L^2} \) {d x \over x} \ln (1 + x)
 \ .
\ee

\ni
The first term is just a numerical constant and the integration over
Grassmann coordinates cancels it. Since the splitting point is
arbitrary we can choose $\xi \gg 1$ and the second term gives

\be
\int d^8 \th \: \ch^{Abel} = {1 \over (4 \pi)^2} \int d^8 \th \: 
 {1 \over 2} \( \ln {W \bar{W} \over \L^2 } \)^2 = {1 \over 4 (4 \pi)^2} 
 \int d^8 \th \, \ln {W^2 \over \L^2} \, \ln {\bar{W}^2 \over \L^2} 
\ee

\ni
to an arbitrary degree of accuracy. In this form it is easy to see
that the IR scale is irrelevant, since all the terms that depend on it
are killed by the $N=2$ superspace integral \cite{n1}, 
\cite{dine_seiberg}. Our calculation gives the correct answer for
the 1-loop nonholomorphic abelian potential with a coefficient

\be
 c = {1 \over 4 (4 \pi)^2} \ .
\ee

\ni
We can alternatively regularize the momentum integral  
using dimensional regularization

\be
 \ch^{Abel} (W, \bar{W}) = {1 \over (4 \pi)^2} \( \mu^2 \)^{-\epsilon}
 \int {d p^2 \over \(p^2\)^{1-\epsilon} } 
 \ln \( 1 + {W \bar{W} \over p^2} \) \ .
\label{my}
\ee

\ni
where $0<\epsilon<1$. Rescaling the momentum variable we 
rewrite (\ref{my})

\be
 \ch^{Abel} (W, \bar{W}) = {1 \over (4 \pi)^2 } 
 \( {W \bar{W} \over \mu^2 } \)^\epsilon 
 \int_{0}^{+\infty} {d y \over y^{1-\epsilon} } \, \ln \, ( 1 + y^{-1} ) \ .
\ee

\ni
The divergence of the integral can be isolated by standard manipulation

\beqs
 \lefteqn{ \int_{0}^{1} {d y \over y^{1-\epsilon} } \( \, \ln ( 1 + y ) -
 \ln y \) \, + \, \int_{1}^{+\infty} {d y \over y^{1-\epsilon} } \,
 \ln ( 1 + y^{-1} )  \nn } \\ \nn \\
& = & \int_{0}^{1} {d y \over y^{1-\epsilon} } \ln ( 1 + y ) \, - \:
 {y^{\epsilon} \over \epsilon} \l[ \ln y - {1 \over \epsilon} \r]_0^1
 \: + \, \int_{0}^{1} {d x \over x^{1+\epsilon} } \ln ( 1 + x ) 
  \nn \\  \nn \\
& = & \int_{0}^{1} {d y \over y} \( y^{-\epsilon} + y ^{\epsilon} \)
 \ln ( 1 + y ) \; - \; {y^{\epsilon} \over \epsilon} \l[ \ln y - 
 {1 \over \epsilon} \r]_0^1    \label{dim_reg} \\
& = & C + {1 \over \epsilon^2} \ .  \nn
\eeqs

 The integral in (\ref{dim_reg}) gives a finite constant and the upper
limit of the last term contains the regulated divergence as we let 
$\epsilon \rightarrow 0$ . The resulting nonholomorphic potential is

\beqs
 \ch^{Abel} (W, \bar{W}) & = & {1 \over (4 \pi)^2 } \:
 lim_{\epsilon \rightarrow 0} \( C + {1 \over \epsilon^2} \) \, 
 \exp \( \epsilon \: \ln {W \bar{W} \over \mu^2 } \)  \nn \\
& = & {1 \over (4 \pi)^2 } \: lim_{\epsilon \rightarrow 0} 
 \( C + {1 \over \epsilon} \ln {W \bar{W} \over \mu^2 } + {1 \over 2} 
 \( \ln {W \bar{W} \over \mu^2 } \)^2 \) \ .
\label{my_div}
\eeqs

The constant $C$ and the chiral divergence $ \ln W + \ln \bar{W}$ are 
killed by the $N=2$ superspace measure and we are left with the same 
nonholomorphic potential we found using an IR cutoff.

\section{Acknowledgments}
We would like to thank R. von Unge and V. Periwal for interesting 
discussions about their work. We acknowledge partial support from NSF 
grant No Phy 9722101.


\begin{thebibliography}{9}

\bibitem{non_hol}{M. Heningson, {\em Nuc. Phys.} {\bf B 458} (1996) 445,
 {\em hep-th}/9507135.} 

\bibitem{n1}{B. de Wit, M.T. Grisaru and M. Ro\v{c}ek, {\em Phys. Lett.}
{\bf B 374} (1996) 297, {\em hep-th}/9601115.}

\bibitem{dine_seiberg}{M. Dine and N. Seiberg, {\em Nuc. Phys.} {\bf B} 
458 (1996) 445, {\em hep-th}/9707057.}

\bibitem{dorey}{N. Dorey, V.V. Khoze, M.P. Mattis, J. Slater and W.A. Weir, 
{\em Phys. Lett.} {\bf B 408} (1997) 213, {\em hep-th}/9706007. \\
D. Bellisai, F. Fucito, M. Matone and G. Travaglini, {\em Phys. Rev.}
{\bf D 56} (1997) 5218, {\em hep-th} /9706099 .}

\bibitem{rikard_vipul}{V. Periwal and R. von Unge, {\em hep-th}/9801099.}

\bibitem{n2_hyper}{F. Gonzalez-Rey, U. Lindstr\"{o}m, M. Ro\v{c}ek, 
R. von Unge and S. Wiles, {\em Nuc. Phys.} {\bf B} in print, 
{\em hep-th}/9710250.\\ 
F. Gonzalez-Rey and R. von Unge, {\em Nuc. Phys.} {\bf B} in print,
{\em hep-th}/9711135.}  

\bibitem{n2_gauge}{F. Gonzalez-Rey, {\em hep-th}/9712128.} 

\bibitem{book}{J. Gates, M. Grisaru, M. Ro\v{c}ek and W. Siegel 
{\em Superspace} Benjamin/Cummings 1983.}

\bibitem{gal}{A. Galperin, E. Ivanov, S. Kalitzin, V. Ogievetsky and E. 
Sokatchev, {\em Class. Quantum Grav.}{\bf 1} (1984) 469,}\\
{A. Galperin, E. Ivanov, V. Ogievetsky and E. Sokatchev,{\em Class. 
Quantum Grav.}{\bf 2} (1985) 601,617.}

\bibitem{disagree}{I.L. Buchbinder, E.I. Buchbinder, S.M. Kuzenko and 
B.A. Ovrut, {\em Phys. Lett.} {\bf B 417} (1998) 61, {\em hep-th}/9704214.\\ 
S. V. Ketov , {\em Phys. Rev.} {\bf D 57} (1998) 1277, {\em hep-th}/9706079.}

\bibitem{murayama}{N. Arkani-Hamed and H. Murayama, 
{\em hep-th}/9705189, {\em hep-th}/9707133.} 





\end{thebibliography}
\end{document}